\documentclass[runningheads]{llncs}

\usepackage[T1]{fontenc}
\def\doi#1{\href{https://doi.org/\detokenize{#1}}{\url{https://doi.org/\detokenize{#1}}}}
\usepackage{graphicx}
\usepackage{listings}
\newcommand\blfootnote[1]{%
  \begingroup
  \renewcommand\thefootnote{}\footnote{#1}%
  \addtocounter{footnote}{-1}%
  \endgroup
}

\begin{document}
%
% \title{A pathologist-informed pipeline for classification of prostate glands in histopathology images\thanks{Supported by organization x.}}
%
\title{A Pathologist-Informed Workflow for Classification of Prostate Glands in Histopathology}


\author{Alessandro Ferrero\inst{1, *} \and
Beatrice Knudsen\inst{2, \dagger} \and
Deepika Sirohi\inst{2, \ddagger}\and \\
Ross Whitaker\inst{1, **}}

%index{Ferrero, Alessandro}
%index{Knudsen, Beatrice}
%index{Sirohi, Deepika}
%index{Whitaker, Ross}

\institute{Scientific Computing and Imaging Institute, University of Utah \\
72 S Central Campus Drive, Room 3750, Salt Lake City, UT 84112 \and
University of Utah \\
201 President Circle, Salt Lake City, UT 84112\\
\email{\inst{*}alessandro.ferrero@utah.edu, \inst{\dagger}Beatrice.Knudsen@path.utah.edu,\\ \inst{\ddagger}Deepika.Sirohi@hsc.utah.edu, \inst{**}whitaker@cs.utah.edu}}

\authorrunning{A. Ferrero et al.}

\maketitle              % typeset the header of the contribution
%
% \begin{abstract}
% Pathologists diagnose and grade prostate cancer by examining tissue from needle biopsies on glass slides. The cancer's severity and risk of metastasis are determined by the Gleason grade, a score based on the organization and morphology of prostate cancer glands. For diagnostic work-up, pathologists first locate glands in the whole biopsy core, and---if they detect cancer---they assign a Gleason grade. This time-consuming process is subject to errors and significant inter-observer variability, despite strict diagnostic criteria. Many machine learning algorithms have been developed to find and grade prostate cancer in whole slide images (WSI) of biopsy tissues, but none of them follows the standard protocol of assessing individual glands. This paper proposes an automated workflow that follows pathologists' \textit{modus operandi}, isolating and classifying multi-scale patches of individual glands using distinct steps: (1) two fully convolutional networks segment epithelium versus stroma and gland boundaries, respectively; (2) a classifier network separates benign from cancer glands at high magnification; and (3) an additional classifier predicts the grade of each cancer gland at low magnification. Altogether, this process provides a gland-specific approach for prostate cancer grading that we compare against other machine-learning-based grading methods.

% \keywords{Prostate cancer \and microscopy imaging \and segmentation \and classification.}
% \end{abstract}

\begin{abstract}
Pathologists diagnose and grade prostate cancer by examining tissue from needle biopsies on glass slides. The cancer's severity and risk of metastasis are determined by the Gleason grade, a score based on the organization and morphology of prostate cancer glands. For diagnostic work-up, pathologists first locate glands in the whole biopsy core, and---if they detect cancer---they assign a Gleason grade. This time-consuming process is subject to errors and significant inter-observer variability, despite strict diagnostic criteria. This paper proposes an automated workflow that follows pathologists' \textit{modus operandi}, isolating and classifying multi-scale patches of individual glands in whole slide images (WSI) of biopsy tissues using distinct steps: (1) two fully convolutional networks segment epithelium versus stroma and gland boundaries, respectively; (2) a classifier network separates benign from cancer glands at high magnification; and (3) an additional classifier predicts the grade of each cancer gland at low magnification. Altogether, this process provides a gland-specific approach for prostate cancer grading that we compare against other machine-learning-based grading methods.

\end{abstract}

\section{Introduction}

\blfootnote{Published as a workshop paper at MICCAI MOVI 2022} Prostate cancer is the second most common cause of cancer death in men over 65 in the Unites States. A reliable diagnosis of prostate cancer can only be accomplished via a prostate needle biopsy. Pathologists examine the extracted tissue samples through  a microscope and assign Gleason grades to cancerous regions as an indicator of cancer severity.
% {
% 	\begin{figure*}
% 		\includegraphics[width=\textwidth]{../Glands_2.pdf}
% 		% \includegraphics{setgan3.pdf}
% 		\caption{A pathologist's workflow for prostate cancer diagnosis and grading: (A) Illustration of key to diagnose cancer. (B) Illustration of key criteria to distinguish low grade (Gleason Grade 3) and high grade (Gleason Grade 4 and 5) cancer areas. These grades correspond to GP3, GP4, and GP5 in the PANDA challenge \cite{PANDA}. In contrast, GP1 and GP2 (PANDA challenge) are labels for stroma and benign glands, respectively.}
% 		\label{fig:glands}
% 	\end{figure*}
% }

The main rationale for Gleason grading is to predict the risk of cancer progression and metastasis that informs treatment decisions. The Gleason grading system encompasses four grades: Gleason grades 2 and 3 are considered low grade and almost never lead to metastatic progression, while Gleason grades 4 and 5 are high grade and carry a risk of metastatic spread. Within the normal prostate tissue, cells organize in tube-like structures called glands. Pathologists use several morphological features to distinguish between cancerous and benign glands. Non-cancerous (i.e., benign) glands consist of basal and luminal cell layers that make up the wall of the tube. The inside of the tube is referred to as the lumen. In contrast, cancerous glands typically loose the basal cell layer, while cancer cell nuclei enlarge and display prominent nucleoli. In addition, the cancerous gland's luminal edge is straight compared to the undulated edge of benign glands.

While cancer diagnosis relies on the cells' organization and appearance, the Gleason grading scheme uses the growth pattern and structural complexity of glands to score the disease's severity. Cancerous glands with a single lumen are classified as low-grade cancer; glands within glands with multiple lumina, or glands that have lost the ability to form a lumen, are high-grade glands. 
% Furthermore, Gleason grade 5 grows as sheets of cells or as single cells invading the surrounding stroma.

Building on deep learning successes in image classification and segmentation, researchers all over the world have turned to neural networks to develop Gleason grading algorithms. Fully convolutional networks \cite{FCN}, in particular, have proven useful in a variety of medical image analysis settings. Typically, a network is trained to recognize and classify structures of interest in the input image, \emph{producing pixel-wise probability maps}, one per class, with each pixel assigned to the class with the highest probability. For instance, Silva-Rodríguez et al. \cite{segmentation1} show that segmenting tumor areas through a neural network achieves better results than traditional algorithms, such as \cite{BCD} and \cite{BCD2}. The U-net (\cite{unet}) is a special form of convolutional-neural-network architecture designed for image segmentation, and many Gleason grading methods such as \cite{bulten2019epithelium,Nagpal,Unet_Corey,Nathan} rely on variants of the U-net to process patches of a whole slide image (WSI) in order to produce a pixel-wise Gleason grade classification. Avinash et al. \cite{Carcinonet} designed their {\em Carcino-net's} architecture to include a pyramid pooling module \cite{PPM} that employs different size convolutional kernels. They show their algorithm's high accuracy on low resolution images that include large cancer areas.  However, this approach does not explicitly account for the gland-level patterns that define the pathology, and, as we will show in later sections, 
Carcino-net may arrive at incorrect conclusions on the gland's grade, even after summarizing pixel-level classification results. Other studies employ region-based convolutional neural networks (RCNNs) \cite{rcnn,maskrcnn} to first identify bounding boxes around areas of interest in a prostate biopsy, and then segment and classify the epithelium within the boxes. The method in \cite{pathrcnn} demonstrates that these RCNNs can identify gland clusters, but struggles when glands with different grades are packed within cancer regions.

This paper aims to accurately reproduce the pathologist’s grading process, breaking the gland classification problem into three sequential tasks: the segmentation of single glands, the identification of malignant glands based on cellular structure, and the classification of glands into low- and high-grade cancer based on the complexity of glandular morphology. In particular, the cancer identification step employs a novel set-based neural network that processes large collections of image patches, summarizing the information into histograms, to distinguish between benign and cancer glands. This \textit{Histogram-based (HB)} workflow provides high gland segmentation accuracy with limited training data. It also allows clinicians and engineers to examine the results at every step of the analysis process. A self-supervised strategy \cite{SSL1,SSL2,SSL3} utilizes nuclear-staining properties to allow better generalization.

\section{Data}

The {\em training dataset}, described in \cite{pathrcnn}, encompasses more than $40,000$ glands, roughly equally distributed among the three classes of interest. The $2,200$ tiles, of $1200 \times 1200$ pixels, that contain the glands were extracted from whole slide images (WSI) at magnification $20$X, with a pixel size of $0.5 \mu m \times 0.5 \mu m$. Several pathologists hand-annotated polygons and assigned a label to the gland outline, marking benign glands, and low-grade (GG3) or high-grade (GG4, GG5) cancer glands. The stroma (ST) between glands is considered background. Through the same process, $10,000$ additional glands from the same forty-one patients were gathered and labeled to form the \emph{Internal test set} ($537$ tiles in total). Annotations mostly corresponded to glands, but clusters of small glands were often included in one outline, making an accurate segmentation difficult to learn. Therefore, $6100$ polygons from the training set were later refined to separate all glands.

% The {\em training dataset}, described in ***ANONYMOUS PAPER***, contains ~$40,000$ glands from 41 patients at ***ANONYMOUS LOCATION***. The magnification is $20$X, with a pixel size of $0.5 \mu m \times 0.5 \mu m$. The dataset includes $2,200$ tiles extracted from the whole slide images (WSI), each with dimensions $1200 \times 1200$. Under supervision of a sub-specialty trained GU  pathologist, several junior pathologists hand-annotated polygons and assigned a label to the gland outline, marking benign glands, and low-grade (GG3) or high-grade (GG4, GG5) cancer glands. The stroma (ST) in between glands is considered background. In the same way, $537$ additional tiles from the same patients were gathered and labeled to form the \emph{in-patient test set} (I-P). Annotations mostly corresponded to glands, but clusters of small glands were often included in one outline within the polygon, creating difficulties in training an algorithm to precisely segment glands. Therefore, $200$ label images from the training set were later refined to separate all glands.

Fully testing the performance of ML-based histology-analysis algorithms requires generalization to data from \textit{unseen patients}. As such, we created the \emph{External test set} by selecting $14800$ glands from WSIs in The Cancer Genome Atlas Program (TCGA). Two pathologists labeled the images from eighteen patients, initially at low resolution to identify regions of different Gleason grades. The $546$ tiles extracted from these polygons were annotated a second time at high resolution, balancing the amount of tiles coming from each class.

% Another set, the \emph{External test set (TCGA)}, was created by selecting $18$ WSIs from The Cancer Genome Atlas Program (TCGA). Two pathologists labeled the images, initially at low resolution to identify regions of different Gleason grades. The $546$ tiles extracted from these polygons were annotated a second time at high resolution, resulting in a set containing $14261$ glands.

For data augmentation, we used random rotation, flipping, and additive noise. As in \cite{hist_match}, color augmentation is performed through histogram matching using  color palettes from TCGA and PANDA {\cite{PANDA}} datasets as target color ranges.

{
	\begin{figure*}
		\includegraphics[width=\textwidth]{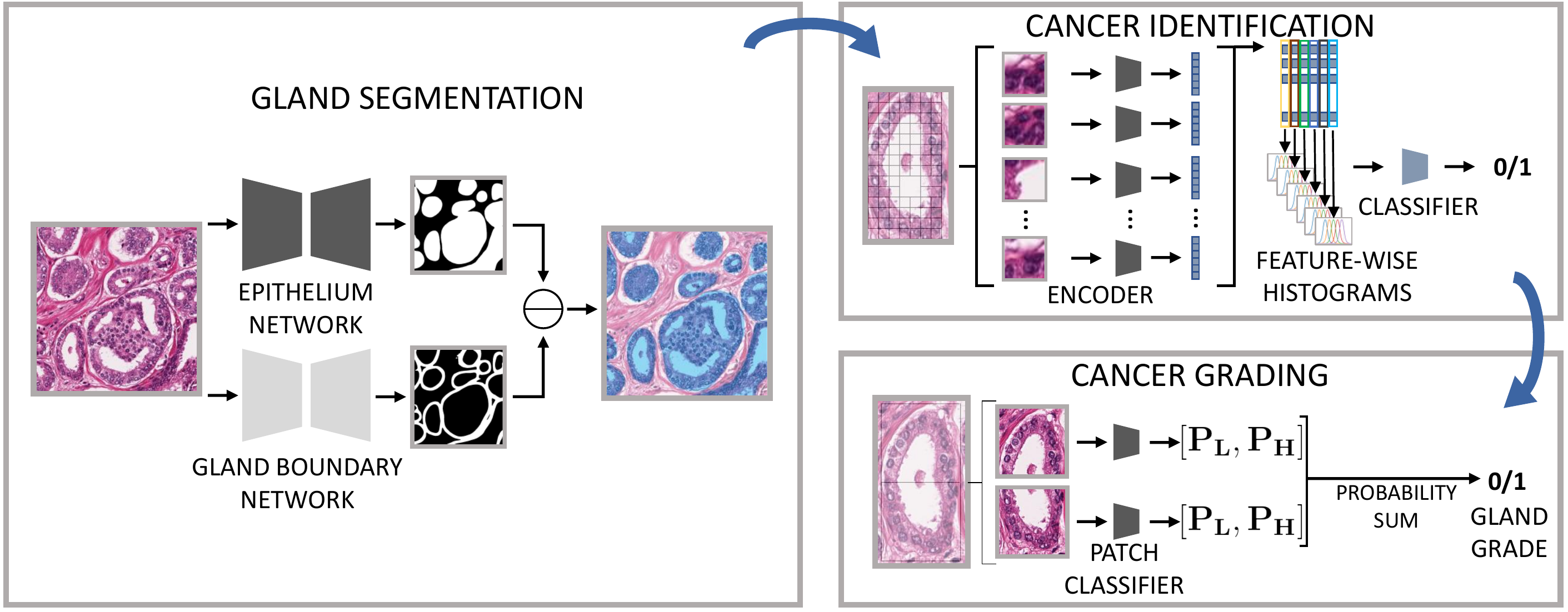}
		\caption{\textit{HB}-workflow. After segmentation, each gland is divided and processed in 32x32 pixel tiles by the cancer identification encoder. Features (n=128) from each gland are placed into 128 histogram bins, which are used to classify between benign and cancerous glands. Next, 64x64 cancer gland pixel tiles are used to distinguish low- and high-grade cancer tiles ($P_L, P_H$). If a gland spans over multiple tiles, the sum of prediction results over all the tiles, normalized to the gland area, determines the gland's predicted grade.}
		\label{fig:architecture}
	\end{figure*}
}

\section{Methods}
\subsection{\textit{HB}-workflow}
The paper's workflow consists of three sequential stages: gland segmentation, cancer gland detection, and cancer grading.

\paragraph{Gland segmentation and processing.} The close proximity of glands within the stroma presents a challenge in separating individual glands. To improve the segmentation, we propose a process that (1) performs epithelium (vs. stroma) segmentation, (2) finds the boundary of glands, (3) identifies the gland lumen to form individual connected components, and (4) expands the connected components to the stromal boundaries to identify entire, distinct glands.
Two identically structured U-net-like architectures perform the first two steps: one network recognizes the epithelium, and the second network finds boundaries around glands in input tiles. As shown by \cite{skip_connections}, short connections between layers increase the prediction accuracy when processing medical images; therefore we employ Resnet blocks \cite{resnet} to capture relevant features. During the training phase, random $256 \times 256$ patches are extracted from the training tiles and the networks learn to minimize the cross-entropy loss between their predictions and the ground truth. The initial learning rate, set at $10^{-4}$, decays every $10$ epochs until the training stops at $2,000$ epochs. Pixel-wise subtraction of the epithelium and the gland boundary reveals the glands as individual, connected components. After removing small components, a region-growing algorithm ensures that gland instances include the entire epithelium.

\begin{table}
\centering
\caption{$mAP_{[0.5, 0.9]}$ on the gland segmentation}\label{ap_table}
\begin{tabular}{|c|c|}
\hline
Internal test set & External test set (TCGA) \\
\hline
$0.67 \pm 0.02$ & $0.77 \pm 0.03$ \\
\hline
\end{tabular}
\end{table}

\paragraph{Cancer gland identification.} Once the individual glands have been segmented, the next stage evaluates the fine-scale cell structure to separate benign and cancerous glands. Since glands may vary widely in size and shape, we design a neural network that accepts sets of image patches from each gland, and outputs a probability of cancer for the entire gland. To construct an appropriate set for analysis, each gland component is divided into $32 \times 32$ overlapping patches, each containing only a few luminal cells that span the thickness of the epithelium between the stroma and the lumen. The network is designed to extract useful cell features from each patch and properly aggregate that information, regardless of the gland size, helping the classifier output a probability of cancer.

\begin{table}
\centering
\caption{Results for CANCER IDENTIFICATION.}\label{results_benign_vs_cancer}
\begin{tabular}{|c|c|c|c||c|c|c|}
\multicolumn{1}{c}{} & \multicolumn{6}{c}{\bf{Internal test set}}\\
\cline{2-7} \multicolumn{1}{c}{} & \multicolumn{3}{|c||}{\bf{Pixel-wise}} & \multicolumn{3}{|c|}{\bf{Gland-wise}}\\
\hline
 & \bf{F1} & \bf{Sensitivity} & \bf{Specificity} & \bf{F1} & \bf{Sensitivity} & \bf{Specificity}\\
\hline
ST & $0.95 \pm 0.01$ & $0.96 \pm 0.01$ & $0.92 \pm 0.01$ & N/A & N/A & N/A\\
BN & $0.89 \pm 0.01$ & $0.91 \pm 0.02$ & $0.97 \pm 0.01$ & $0.95 \pm 0.01$ & $0.95 \pm 0.01$ & $0.95 \pm 0.02$\\
CN & $0.89 \pm 0.01$ & $0.86 \pm 0.01$ & $0.98 \pm 0.01$ & $0.96 \pm 0.01$ & $0.93 \pm 0.02$ & $0.99 \pm 0.01$\\
\hline
\end{tabular}

\begin{tabular}{c c }
\centering
 & 
\end{tabular}
\begin{tabular}{|c|c|c|c||c|c|c|}
\multicolumn{1}{c}{} & \multicolumn{6}{c|}{\bf{External test set}}\\
\cline{2-7} \multicolumn{1}{c}{} & \multicolumn{3}{|c||}{\bf{Pixel-wise}} & \multicolumn{3}{|c|}{\bf{Gland-wise}}\\

\hline
 & \bf{F1} & \bf{Sensitivity} & \bf{Specificity} & \bf{F1} & \bf{Sensitivity} & \bf{Specificity}\\
\hline
ST & $0.97 \pm 0.01$ & $0.94 \pm 0.01$ & $0.98 \pm 0.01$ & N/A & N/A & N/A\\
BN & $0.74 \pm 0.04$ & $0.80 \pm 0.06$ & $0.93 \pm 0.03$ & $0.76 \pm 0.04$ & $0.81 \pm 0.03$ & $0.89 \pm 0.06$\\
CN & $0.89 \pm 0.01$ & $0.88 \pm 0.04$ & $0.94 \pm 0.01$ & $0.96 \pm 0.01$ & $0.88 \pm 0.04$ & $0.83 \pm 0.06$\\
\hline
\end{tabular}

\begin{tabular}{c c }
\centering
 & 
\end{tabular}
\end{table}

Inspired by the SetGAN discriminator's design in \cite{Ferrero2019SetGANIT}, the proposed NN architecture includes three modules: (1) an encoder that processes patches individually and generates a 128-dimensional feature vector; (2) a surrogate histogram function that summarizes each feature along all patches from a gland into a histogram with $k$ bins (obtaining one histogram per feature), and (3) fully connected layers that use the resulting 128 histograms to output the gland classification. In this work, $k$ is set to $5$: empirically, fewer that 5 bins lead to an insufficiently descriptive latent representation, while more than $5$ bins do not provide much additional information.  This architecture has two advantages over a CNN: while the aggregating histogram function provides a rich representation of the whole gland, regardless of its size, the permutation invariance peculiar to the design (i.e., the patch order does not affect the classification) allows the network to study small cell groups, regardless of their location within the gland.

\paragraph{Cancer gland grading.} The final step consists of classifying cancer glands into high or low grades. An analysis of the morphology of the entire gland, including its lumen, is necessary for this task. The above approach, developed for cancer detection, is not expected to work for cancer grading, since the unordered collection of small patches does not capture the complex morphology of high-grade cancer glands.  Furthermore, Ma et al. \cite{Nathan} show that analyzing glands at a lower magnification yields better results when assigning Gleason grades. In order to format each malignant gland from the previous stage into a more appropriate set of NN inputs, we reduce the magnification to $10$X via a down sampling by a factor of $2$ and then use $64 \times 64$ patches.  A Resnet architecture learns to assign each cancer patch as high- and low-grade class.  
While most glands can be contained in a single patch, glands that span multiple patches are graded based on a majority vote.
{
	\begin{figure*}
	    \centering
		\includegraphics[scale=0.43]{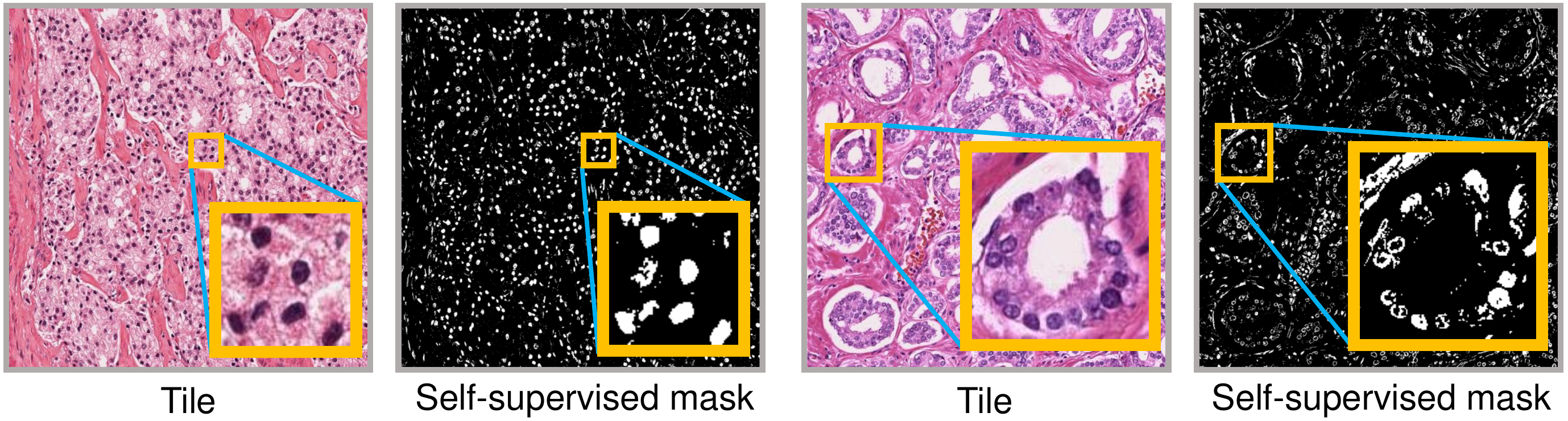}
		\caption{Masks generated by thresholding the $10\%$ darkest pixels.}
		\label{fig:augmentation}
	\end{figure*}
}
% \caption{Masks generated by thresholding the $10\%$ darkest pixels. Most pixels are seen within nuclear boundaries. The low magnification image demonstrates difference in nuclear organization between high grade (upper panel) and low grade (lower panel) prostate cancer. The high magnification inserts demonstrate additional resolution of nuclear characteristics related to increased nuclear size and prominence of nucleoli.

\subsection{Self-supervised strategy}

% To further improve generalization, we employ a self-supervised, multi-task strategy throughout the entire pipeline: all encoders train in association with a decoder that learns to segment nuclei in input images.
% The hematoxylin stain gives cell nuclei a purple/blue hue---usually the darkest color in prostate WSIs---therefore thresholding the darkest $10\%$ pixels in the training tiles creates the labels for the task. Despite some noise, Figure \ref{fig:augmentation} shows that the cell nuclei are discernible within the image. This self-supervised task forces encoders to learn general cell features, useful for segmenting and classifying glands.

To further improve generalization, we employ a self-supervised strategy throughout the entire pipeline. When prostate tissue is stained with hematoxylin and eosin (H\&E), cell nuclei acquire a purple/blue hue, usually the darkest color in prostate WSIs. The self-supervised task labels are obtained by thresholding the $10\%$ darkest training tile pixels (see Figure \ref{fig:augmentation}). The resulting mask highlights cell nuclei. All encoders train in association with a decoder that learns this coarse nuclear segmentation, encouraging encoders to learn nuclear features useful for segmenting and classifying glands. Altogether, our self-supervised approach is motivated by pathologist-defined diagnostic cues related to nuclear features.

\begin{table}
\centering
\caption{F1 scores on the \emph{Internal test set}, similar to the training set.}\label{results_internal}
\begin{tabular}{|c|c|c|c||c|c|c|}
\cline{2-7} \multicolumn{1}{c}{} & \multicolumn{3}{|c||}{\bf{Pixel-wise F1 scores}} & \multicolumn{3}{|c|}{\bf{Gland-wise F1 scores}}\\
\hline
Class &  U-net & Carcino-net & 3-stage & U-net & Carcino-net & 3-stage\\
\hline
ST & $0.92 \pm 0.01$ & $\mathbf{0.95 \pm 0.01}$ & $0.95 \pm 0.01$ & N/A & N/A & N/A\\
BN & $0.84 \pm 0.02$ & $\mathbf{0.93 \pm 0.01}$ & $0.89 \pm 0.01$ & $0.95 \pm 0.01$ & $\mathbf{0.97 \pm 0.01}$ & $0.94 \pm 0.01$\\
LG & $0.52 \pm 0.02$ & $\mathbf{0.79 \pm 0.01}$ & $0.73 \pm 0.01$ & $0.71 \pm 0.01$ & $\mathbf{0.91 \pm 0.01}$ & $0.86 \pm 0.02$\\
HG & $0.68 \pm 0.02$ & $\mathbf{0.89 \pm 0.02}$ & $0.86 \pm 0.01$ & $0.80 \pm 0.01$ & $\mathbf{0.94 \pm 0.01}$ & $0.93 \pm 0.01$\\
\hline
\cline{2-7} \multicolumn{1}{c}{} & \multicolumn{3}{|c||}{\bf{Pixel-wise Sensitivity}} & \multicolumn{3}{|c|}{\bf{Gland-wise Sensitivity}}\\
\hline
Class &  U-net & Carcino-net & 3-stage & U-net & Carcino-net & 3-stage\\
\hline
ST & $0.93 \pm 0.01$ & $\mathbf{0.94 \pm 0.01}$ & $0.96 \pm 0.01$ & N/A & N/A & N/A\\
BN & $0.80 \pm 0.02$ & $\mathbf{0.94 \pm 0.01}$ & $0.91 \pm 0.01$ & $0.94 \pm 0.01$ & $\mathbf{0.97 \pm 0.01}$ & $0.95 \pm 0.02$\\
LG & $0.81 \pm 0.02$ & $\mathbf{0.76 \pm 0.01}$ & $0.75 \pm 0.03$ & $0.95 \pm 0.01$ & $\mathbf{0.86 \pm 0.01}$ & $0.84 \pm 0.02$\\
HG & $0.53 \pm 0.02$ & $\mathbf{0.91 \pm 0.01}$ & $0.81 \pm 0.02$ & $0.67 \pm 0.01$ & $\mathbf{0.96 \pm 0.01}$ & $0.90 \pm 0.01$\\
\hline
\cline{2-7} \multicolumn{1}{c}{} & \multicolumn{3}{|c||}{\bf{Pixel-wise Specificity}} & \multicolumn{3}{|c|}{\bf{Gland-wise Specificity}}\\

\hline
ST & $0.89 \pm 0.01$ & $\mathbf{0.93 \pm 0.01}$ & $0.92 \pm 0.01$ & N/A & N/A & N/A\\
BN & $0.97 \pm 0.01$ & $\mathbf{0.97 \pm 0.01}$ & $0.95\pm 0.02$ & $0.98 \pm 0.01$ & $\mathbf{0.98 \pm 0.01}$ & $0.95 \pm 0.02$\\
LG & $0.90 \pm 0.01$ & $\mathbf{0.98 \pm 0.01}$ & $0.97\pm 0.01$ & $0.86 \pm 0.01$ & $\mathbf{0.99 \pm 0.01}$ & $0.98 \pm 0.01$\\
HG & $0.98 \pm 0.01$ & $\mathbf{0.96 \pm 0.01}$ & $0.98\pm 0.01$ & $0.99 \pm 0.01$ & $\mathbf{0.98 \pm 0.01}$ & $0.98 \pm 0.01$\\
\hline
\end{tabular}
\end{table}

\section{Results}
To quantitatively evaluate the model, the \textit{HB}-workflow trained end-to-end ten times, using random validation sets of $2400$ glands from the \emph{External test set} and testing on the remaining samples. F1 scores, sensitivity and specificity are calculated per class (stroma - ST, benign - BN, low-grade - LG, high-grade - HG) to evaluate the pixel-wise and gland-wise performance. The \textit{HB}-workflow's pixel-level data were generated using the gland-level classification labels. Gland-wise scores are obtained by majority vote of the corresponding pixels in the prediction mask, with the final score weight proportional to the gland's size.

{
	\begin{figure*}
		\includegraphics[width=\textwidth]{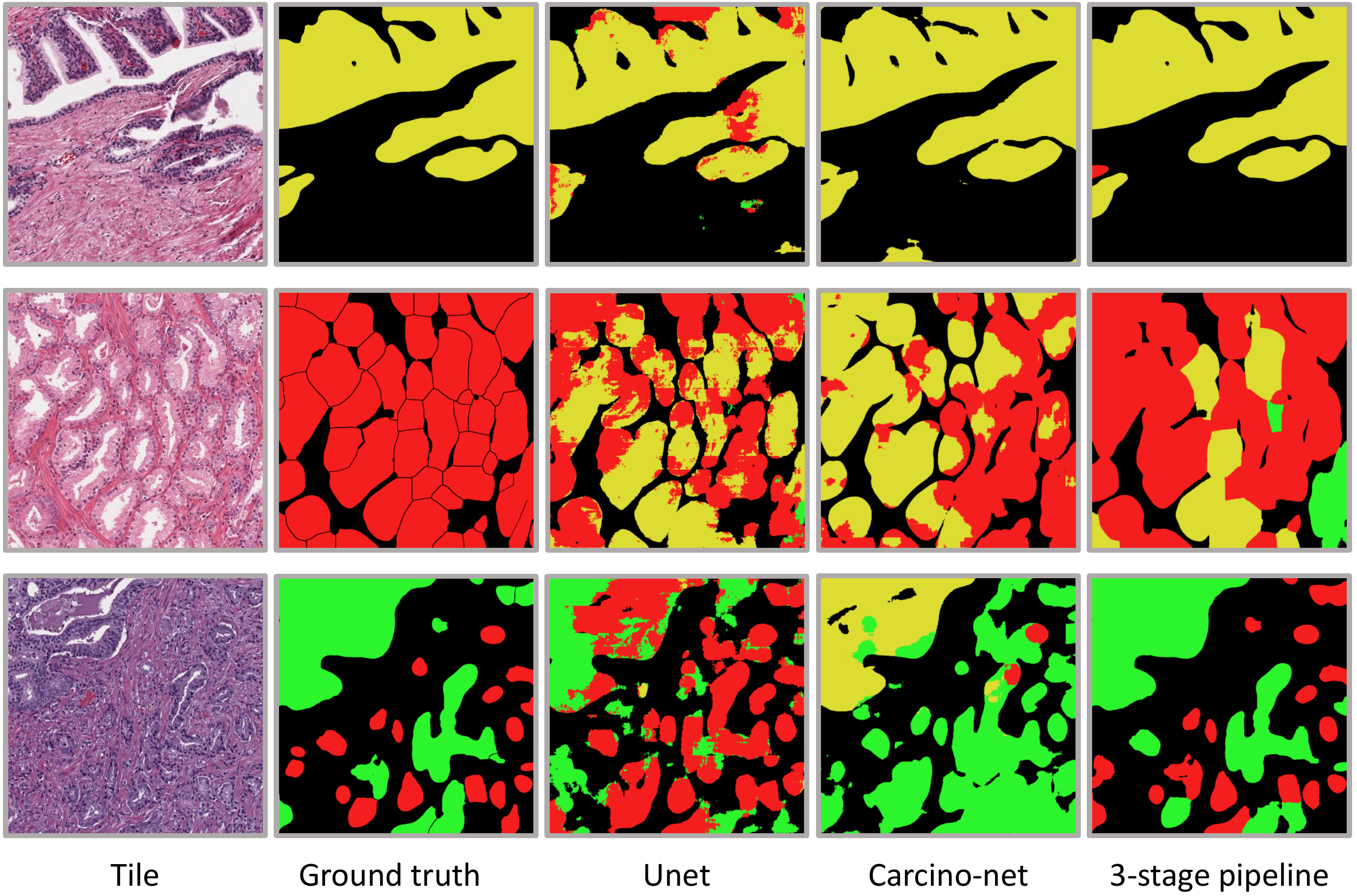}
		\caption{Classification results. H\&E stained image tiles with manual ground truth labels are compared to U-net and Carcino-net pixel-level classification results and gland level classification from the \textit{HB}-workflow (yellow - BN, red - LG, green - HG).  }
		\label{fig:result_imgs}
	\end{figure*}
	\label{prediction_imgs}
}

Table \ref{ap_table} shows the mean average precision (mAP) between the manually segmented glands and the predicted glands. Although the \textit{HB-workflow} tends to slightly oversegment glands (especially when the high-grade cancer consists of free cells), the mAP values are high. Experiments showed that gland oversegmentation is less detrimental to the final workflow's output than undersegmentation, where the binary classifiers tend to give class probabilities closer to $0.5$ when glands of different grades appear in the same segmented instance (i.e., the classifiers are less certain about their predictions and make more mistakes). The slight difference between the datasets' scores in Table \ref{ap_table} is mainly due to the gland conglomerates present in the \textit{Internal test set}'s labels that decrease the mAP value. Despite this, the mAP values remain high.

Evaluations in Table \ref{results_benign_vs_cancer} show that the cancer-identifying classifier distinguishes between benign and cancer glands with high accuracy, thanks to the rich representation provided by the learned histograms. Tables \ref{results_internal} and \ref{results_external} compare U-net and Carcino-Net, trained as described in \cite{Carcinonet}, to the \textit{HB}-workflow.

Figure \ref{fig:result_imgs} confirms the high scores that the three architectures achieve in segmenting stroma. Panels displaying the U-net and Carcino-net pixel-level classification results demonstrate misclassified pixels within glands. These pixel labels are corrected by applying the majority voting scheme, resulting in a single label for each gland. The quantitative results in Tables \ref{results_internal} and \ref{results_external} show that Carcino-Net performs better on the \textit{Internal test set} (which is similar to the training data) than on the \textit{External test set}, meaning that Carcino-Net is prone to overfitting. In contrast, the \textit{HB}-workflow achieves higher F1 scores, pixel-wise and gland-wise, for all classes on the unseen data.

\begin{table}
\centering
\caption{Results on the \emph{External test set (TCGA)}. }\label{results_external}
\begin{tabular}{|c|c|c|c||c|c|c|}
\cline{2-7} \multicolumn{1}{c}{} & \multicolumn{3}{|c||}{\bf{Pixel-wise F1 scores}} & \multicolumn{3}{|c|}{\bf{Gland-wise F1 scores}}\\
\hline
Class &  U-net & Carcino-net & \textit{HB workflow} & U-net & Carcino-net & \textit{HB workflow}\\
\hline
ST & $0.89 \pm 0.01$ & $0.91 \pm 0.01$ & $\mathbf{0.97 \pm 0.01}$ & N/A & N/A & N/A\\
BN & $0.58 \pm 0.02$ & $0.72 \pm 0.11$ & $\mathbf{0.74 \pm 0.04}$ & $0.69 \pm 0.05$ & $0.74 \pm 0.06$ & $\mathbf{0.76 \pm 0.04}$\\
LG & $0.58 \pm 0.04$ & $0.66 \pm 0.03$ & $\mathbf{0.69 \pm 0.03}$ & $0.67 \pm 0.07$ & $0.68 \pm 0.08$ & $\mathbf{0.71 \pm 0.04}$\\
HG & $0.39 \pm 0.14$ & $0.55 \pm 0.06$ & $\mathbf{0.69 \pm 0.10}$ & $0.54 \pm 0.18$ & $0.68 \pm 0.09$ & $\mathbf{0.72 \pm 0.01}$\\
\hline
\end{tabular}
\begin{tabular}{|c|c|c|c||c|c|c|}
\cline{2-7} \multicolumn{1}{c}{} & \multicolumn{3}{|c||}{\bf{Pixel-wise Sensitivity}} & \multicolumn{3}{|c|}{\bf{Gland-wise Sensitivity}}\\
\hline
Class &  U-net & Carcino-net & \textit{HB workflow} & U-net & Carcino-net & \textit{HB workflow}\\
\hline
ST & $0.85 \pm 0.01$ & $0.87 \pm 0.01$ & $\mathbf{0.94 \pm 0.01}$ & N/A & N/A & N/A\\
BN & $0.52 \pm 0.02$ & $\mathbf{0.87 \pm 0.01}$ & $0.80 \pm 0.06$ & $0.55 \pm 0.01$ & $\mathbf{0.88 \pm 0.01}$ & $0.81 \pm 0.03$\\
LG & $\mathbf{0.83 \pm 0.03}$ & $0.62 \pm 0.01$ & $0.73 \pm 0.02$ & $\mathbf{0.90 \pm 0.01}$ & $0.64 \pm 0.01$ & $0.73 \pm 0.02$\\
HG & $0.36 \pm 0.14$ & $0.58 \pm 0.01$ & $\mathbf{0.65 \pm 0.04}$ & $0.41 \pm 0.01$ & $0.63 \pm 0.01$ & $\mathbf{0.66 \pm 0.01}$\\
\hline
\end{tabular}
\begin{tabular}{|c|c|c|c||c|c|c|}
\cline{2-7} \multicolumn{1}{c}{} & \multicolumn{3}{|c||}{\bf{Pixel-wise Specificity}} & \multicolumn{3}{|c|}{\bf{Gland-wise Specificity}}\\
\hline
Class &  U-net & Carcino-net & \textit{HB workflow} & U-net & Carcino-net & \textit{HB workflow}\\
\hline
ST & $0.93 \pm 0.01$ & $0.93 \pm 0.01$ & $\mathbf{0.98 \pm 0.01}$ & N/A & N/A & N/A\\
BN & $\mathbf{0.98 \pm 0.02}$ & $0.89 \pm 0.01$ & $0.93\pm 0.03$ & $\mathbf{0.98 \pm 0.01}$ & $0.86 \pm 0.01$ & $0.89 \pm 0.06$\\
LG & $0.80 \pm 0.04$ & $\mathbf{0.93 \pm 0.0}1$ & $0.90\pm 0.01$ & $0.53 \pm 0.01$ & $\mathbf{0.88 \pm 0.01}$ & $0.84 \pm 0.04$\\
HG & $\mathbf{0.94 \pm 0.05}$ & $0.89 \pm 0.01$ & $0.89\pm 0.01$ & $\mathbf{0.94 \pm 0.01}$ & $0.86 \pm 0.01$ & $0.86 \pm 0.06$\\
\hline
\end{tabular}
\end{table}

\section{Conclusions}
Pathologists diagnose and grade prostate cancer glands based on vastly different morphological criteria. The \textit{HB}-workflow presented in this paper mimics the pathologist's workflow by separating gland segmentation, cancer detection and cancer grading into three separate stages. The division of tasks allows each neural network to focus on the relevant features for the task at hand. In particular, the histogram aggregation function provides a permutation invariant way to process sets of small gland patches, allowing the cancer identification network to focus on the cell morphology. The \textit{HB}-workflow shows higher quantitative and qualitative results per class than other state-of-the-art methods. Future work includes the training of this pipeline on larger, multi-cohort datasets, and its use for identifying high grade cancer regions to cost-effectively predict cancer stage and prognosis.

\bigskip

\noindent
\textbf{Acnowledgments.} We acknowledge the generous support from the Department of Defense Prostate Cancer Program Population Science Award W81XWH-21-1-0725-. We also acknowledge that we received the training data from Cedars-Sinai Hospital in Los Angeles and we thank Dr. Akadiusz Gertych for his work on establishing the tiles. The results presented here are in part based upon data generated by the TCGA Research Network: https://www.cancer.gov/tcga.

\bibliographystyle{splncs04}
\bibliography{Pathologist_informed_workflow}

\end{document}